\documentclass{article}

\usepackage{microtype}
\usepackage{graphicx}
\usepackage{subfigure}
\usepackage{subcaption}
\usepackage{booktabs} 
\usepackage{lipsum}
\usepackage[dvipsnames]{xcolor}
\usepackage[hyphens]{url}
\usepackage{xurl}
\usepackage{xcolor}

\usepackage{hyperref}

\usepackage{tikz}
\usetikzlibrary{positioning}

\newcommand{\raisedsim}{\raise.17ex\hbox{$\scriptstyle\sim$}}

\usepackage[accepted]{icml2025}

\usepackage{amsmath}
\usepackage{amssymb}
\usepackage{mathtools}
\usepackage{amsthm}

\usepackage[capitalize,noabbrev]{cleveref}

\theoremstyle{plain}

\theoremstyle{definition}

\theoremstyle{remark}

\icmltitlerunning{Distributed and Decentralised Training: Technical Governance Challenges in a Shifting AI Landscape}

\begin{document}

\twocolumn[
\icmltitle{Distributed and Decentralised Training: Technical Governance Challenges in a Shifting AI Landscape}



\icmlsetsymbol{equal}{*}

\begin{icmlauthorlist}
\icmlauthor{Jakub Kryś}{mats}
\icmlauthor{Yashvardhan Sharma}{mats,minerva}
\icmlauthor{Janet Egan}{cnas}
\end{icmlauthorlist}

\icmlaffiliation{mats}{ML Alignment \& Theory Scholar}
\icmlaffiliation{minerva}{Minerva University}
\icmlaffiliation{cnas}{Center for a New American Security}

\icmlcorrespondingauthor{Jakub Kryś}{jakkrys@gmail.com}

\icmlkeywords{Machine Learning, ICML}

\vskip 0.3in
]



\printAffiliationsAndNotice{}  

\begin{abstract}
Advances in low-communication training algorithms are enabling a shift from centralised model training to compute setups that are either distributed across multiple clusters or decentralised via community-driven contributions. This paper distinguishes these two scenarios -- distributed and decentralised training -- which are little understood and often conflated in policy discourse. We discuss how they could impact technical AI governance through an increased risk of compute structuring, capability proliferation, and the erosion of detectability and shutdownability. While these trends foreshadow a possible new paradigm that could challenge key assumptions of compute governance, we emphasise that certain policy levers, like export controls, remain relevant. We also acknowledge potential benefits of decentralised AI, including privacy-preserving training runs that could unlock access to more data, and mitigating harmful power concentration. Our goal is to support more precise policymaking around compute, capability proliferation, and decentralised AI development.
\end{abstract}

\section{Introduction}
\label{sec:intro}
Large Language Model (LLM) training has rapidly grown in scale and now necessitates the use of massive computational resources~\cite{epoch2024trainingcompute}. Historically, this compute has been delivered in a single large data centre housing tightly interconnected GPUs. Such centralised settings are an attractive lever for AI governance as they are easily detectable and concentrated in the hands of a few entities~\cite{sastry2024computing}.

However, recent algorithmic developments indicate that it might be possible to train large-scale AI models on pools of compute that are not located in the same place and that can leverage slower connections. By drastically reducing the inter-GPU communication needs, these developments open the door to two new training paradigms: (1) hyperscalers performing training runs across multiple geographically dispersed clusters and (2) community-driven training of models over the Internet with no centralised point of control. Progress is happening fast on both fronts. Several frontier models have already been trained by combining compute from multiple data centres~\cite{team2024gemini, youtube2025openai}. At the same time, since around the beginning of 2024 there has been a rapid rise of startups that use decentralised approaches to training AI models. We estimate these startups have raised around \$145 million in venture capital funding so far\footnote{This number is the aggregate funding received by the most prominent startups operating in this space: Nous Research -- \$50M~\cite{nous_funding}, Gensyn AI -- \$43M~\cite{gensyn_funding}, Prime Intellect -- \$20M~\cite{prime_funding}, Flower Labs -- \$20M~\cite{flower_funding}, Pluralis Research -- \$7.6M~\cite{pluralis_funding} and Berkeley Compute -- \$5M~\cite{berkeley_funding}.}, with the goal of training o3-level models by the end of 2025~\cite{prime2025math}.

Both scenarios have important implications for the future of technical AI governance and we believe they should occupy more space in the policymaking discourse. As a first step, we offer the following contributions:
\begin{itemize}
    \item We propose that the AI policy community clearly distinguish between the terms `distributed’ and `decentralised’ training to help delineate these scenarios and enable more precise discussion.
    \item We briefly describe the algorithmic progress enabling these developments, as well as their motivations and likely future trajectory.
    \item We detail several implications of this emerging trend on AI governance, including an increased risk of compute structuring and the erosion of detectability and shutdownability.
    \item We highlight possible benefits of decentralised training, including access to more data in a privacy-preserving manner and mitigation of power concentration risks.
\end{itemize}

\begin{figure}
    \centering
    \begin{tikzpicture}[
        every node/.style={font=\sffamily},
        box/.style={minimum width=2.8cm, minimum height=1cm, draw}
    ]
    
    \draw (0,0) rectangle (5.6,3.2);
    \draw (2.8,0) -- (2.8,3.2);
    \draw (0,1.6) -- (5.6,1.6);
    
    \node[rotate=90] at (-0.8,1.6) {Contributing parties};
    \node at (2.8,4.0) {Compute locations};
    
    \node at (1.4,3.5) {One};
    \node at (4.2,3.5) {Many};
    \node[rotate=90] at (-0.3,2.4) {One};
    \node[rotate=90] at (-0.3,0.8) {Many};
    
    \node[align=center] at (1.4,2.4) {centralised \\ {(\textcolor{brown}{Grok 3})}};
    \node[align=center] at (4.2,2.4) {distributed \\ {(\textcolor{blue}{GPT-4.5})}};
    \node[align=center] at (1.4,0.8) {multi-party \\ centralised \\ {\textcolor{red}{\large ?}}};
    \node[align=center] at (4.2,0.8) {decentralised \\ {(\textcolor{ForestGreen}{INTELLECT-1})}};
    \end{tikzpicture}
    \vspace{-0.2em}
    \caption{Possible training regimes organised by the number of contributing parties and compute locations. Each quadrant contains example models: \textcolor{brown}{Grok 3} represents physically centralised compute with a single coordinating party~\cite{grok3}; \textcolor{blue}{GPT-4.5} illustrates centralised coordination across many compute sites~\cite{gpt45_syscard}; \textcolor{ForestGreen}{INTELLECT-1} exemplifies decentralised training with many contributors and locations~\cite{jaghouar2024intellect}; and \textcolor{red}{?} marks the hypothetical scenario of multiple parties contributing compute placed in one location.}
    \label{fig:grid}
\end{figure}
\section{Distributed and Decentralised Training} \label{sec:distributed}
\subsection{Distinction}
We believe it is useful to clearly distinguish between two emerging paradigms: distributed and decentralised training (see Fig.~\ref{fig:grid}). Historically, distributed training was a purely technical term referring to training models on several GPUs. Today, due to model size, every training run of frontier LLMs is distributed across many GPUs, rendering the original meaning obsolete. Thus, we propose the following terminology:
\begin{itemize}
    \item \textbf{Distributed training} -- training across multiple physically distant pools of compute, with a central entity coordinating the process. Other appropriate names could be `multi-data centre training' or `geographically distributed training'.
    \item \textbf{Decentralised training} -- training that involves community-provided compute pools and contributions without a central coordinating entity.
\end{itemize}
Both paradigms leverage the same technical advances but differ in motivations and AI governance implications, warranting a clear distinction between them\footnote{At present, these terms are often used interchangeably~\cite{bye2025nightmare}, yet they pose distinct challenges. The case for making this distinction has also been made by~\cite{lehman2025symbolic, knower2025}.}.
\subsection{Distributed training}
Building clusters large enough for frontier LLM training poses significant challenges, from permitting to accessing the energy needed to power them. Power demands for the largest training runs today are well above 100 megawatts – equivalent to supplying around 80,000 households with electricity – and are projected to grow to 5 gigawatts or above by 2030~\cite{epoch2024canaiscaling, ifp_compute, pilz2025trendsaisupercomputers}. Given the difficulty of building and accessing energy at this scale in a single location, an alternative is to use multiple smaller data centres placed in different areas. Several AI labs already use multi-data centre training, for example in training GPT-4.5 and Gemini-1.5, although the exact distances between these data centres are not publicly known~\cite{team2024gemini, youtube2025openai}. We anticipate that power constraints will continue to drive increased adoption of such distributed AI training.

To accomplish this,  training runs can utilise so-called 'data parallelism', which is a method of training several copies of the model concurrently. Each copy receives a subset of the training data, updates its own parameters, then synchronises them with the other copies to produce an optimal model. To perform this synchronisation, GPUs have traditionally been interconnected through a very high bandwidth network, enabled by specialised equipment within a single cluster. Long-distance training is significantly more challenging, but analysis suggests that hyperscalers could achieve sufficient synchronisation by tapping into the vast amounts of unused fibre optic cables  across the United States known as `dark fibre' ~\cite{semianalysis2024multidatacentre, zayo2024darkfibre}. This could enable hyperscalers to train models across much greater distances.

\subsubsection{Relevant algorithmic progress}
New training methods that build on data parallelism can also help enable more ambitious uses of distributed training. Recently developed low-communication algorithms allow for much less intensive synchronisation of the data-parallel copies, often by as much as a factor of 500 (for a brief overview, see Appendix~\ref{app:low-com})~\cite{douillard2023diloco, liu2024asynchronous, douillard2025streaming, kale2025eager, peng2024decoupled}. In practical terms, this potentially enables
conducting training runs over networks as slow as typical Internet speeds. This means that the importance of some types of interconnects could now be significantly lower, aiding efforts to synchronise GPUs less frequently and over longer distances. We will discuss the implications of this effect in Sec.~\ref{sec:implications}.

\subsection{Decentralised Training} \label{sec:decentralised}
This same algorithmic progress in low-communication training is also supporting new approaches to \textit{decentralised} training. Several startups, most notably Prime Intellect and Nous Research, have already managed to pre-train models at the scale of 10 billion parameters on GPUs that were placed on different continents and communicated using typical Internet speeds~\cite{jaghouar2024intellect, nous2024distro15b}. While these models currently underperform on benchmarks relative to similarly-sized counterparts, it is unclear to what extent this should be interpreted as a sign against their ultimate potential. For now, these efforts predominantly focused on demonstrating the feasibility of global decentralised training runs, rather than achieving optimal performance. Further improvements are likely achievable, though the ultimate competitiveness gap remains uncertain\footnote{In particular, recent benchmarking results of reasoning models have questioned the quality of INTELLECT-2, a reasoning model post-trained in a decentralised manner, which seems to underperform relative to its base model~\cite{hochlehnert2025sober}. However, it is unclear whether this effect is due to its decentralised training procedure or other methodological flaws that apply to RL post-training more broadly.}.

On the whole, the overarching goal of decentralised AI is to democratise access to AI, encourage open innovation and offer a counterbalance to the hyperscalers' dominance. The aim of these initiatives is to accumulate enough community-driven compute to enable the creation of collectively-owned, open-source models that can rival leading industry products. Indeed, several precedents for globally pooled compute already exist, for example the Folding@Home project which at its peak accumulated 280 000 GPUs and was the first compute pool to cross the $10^{18}$ FLOPS threshold~\cite{Zimmerman2021}\footnote{The project used around 280 000 GPUs and 4.8 million CPUs at its peak, but after accounting for their relative performance, we deduce that the GPUs were responsible for around 94\% of the total FLOPS.}. However, it should be noted that AI training (and inference) involves very different computational workloads, so this raw FLOPS count should not be directly translated to the AI context. Nonetheless, it illustrates the rough magnitude of decentralised compute that could be accumulated. We leave a precise estimate of the largest possible decentralised training run to future work.

\subsubsection{Peer-to-Peer Communication} \label{sec:p2p}
Unlike the multi-data centre distributed training described in Sec.~\ref{sec:distributed}, decentralised training runs do not rely on a single point of control. Most GPU network layouts involve at least one type of centralisation: (1) algorithmic centralisation, where model synchronisation happens through a single server, or (2) operational centralisation, where an organisational layer coordinates what each GPU should do, monitors their status and manages resources. The opposite design can be implemented in so-called peer-to-peer (P2P) networks. Here, there is no central choke point and all tasks are initiated and coordinated by the network nodes themselves. When training an AI model, new nodes can contribute to the process \textit{trustlessly}, that is without an approval by a central authority. The legitimacy of their work is attested through cryptographic proofs and verified by other nodes~\cite{prime2025toploc}. Similarly, model synchronisation can be done without a master node, instead relying on peers coordinating with other peers directly~\cite{ryabinin2021moshpit}. Even if a given GPU fails, this does not halt the training run -- the network can dynamically reorganise, continue training, and accept replacement GPUs that obtain key instructions and information from their peers~\cite{hivemind, jaghouar2024intellect}.

In terms of design, this setup is much more closely related to the decentralised structure of blockchains such as Ethereum than it is to traditional data-parallel training. While no training protocol has achieved a \textit{fully} decentralised status as of today, this is a stated goal~\cite{prime2025protocol}. This could signal the emergence of a new regime, one where oversight is much harder and some of the key assumptions underpinning compute governance are questioned.

\section{Implications for Governance} \label{sec:implications}
In this section, we discuss the implications of these trends for AI governance. Advances in distributed training will likely enable leading AI companies to bypass energy constraints and continue to scale, reinforcing their dominance at the forefront of AI. Decentralised training, however, comes with a novel set of challenges and opportunities.

\subsection{Capability Proliferation}
Decentralised AI -- alongside distillation, algorithmic progress and hardware efficiency -- could be one of the major drivers of AI capability proliferation~\cite{pilz2025increased}. This trend could be especially significant if traditional open-weight providers shift to closed-weight deployment, curtailing the usual routes for unrestricted access. Moreover, it offers an alternative path for a broader set of actors to contribute their compute (potentially including consumer GPUs) to develop and deploy advanced AI systems. Notably, the post-training of models to unlock stronger capabilities is particularly well-suited to distributed and decentralised environments, which favour techniques that operate effectively under limited communication (see Appendix~\ref{app:reasoning}). We therefore expect that access to, and ownership of, increasingly powerful models will become more commoditised. While it is unclear to what extent this trend will scale, progress since the beginning of 2024 has been impressive and suggests this issue deserves further attention of the AI governance community. We echo calls from other AI researchers to prioritise research on societal resilience~\cite{tonernonprolif}.

Technical feasibility aside, the motivation to make decentralised AI happen is apparent -- we should not be surprised if the next few years lead to a real explosion in uptake of these technologies, similarly to blockchain-style networks a decade ago, and with no clear way of overseeing or regulating them. We do not take a stance on whether decentralising AI development is positive or not, but we call for continued evaluation of its risk-benefit profile through the `marginal risk' framework~\cite{kapoor2024societal}. If decentralised AI significantly lags behind the frontier and sufficient defensive technologies can be developed, government intervention may be unnecessary, especially given the potential benefits of decentralisation.

\subsection{Shutdownability}

Decentralised training could undermine policy interventions that aim to promote ‘shutdownability’ of powerful AI models. To manage potentially catastrophic AI risks, it has been argued that we should preserve an option of shutting down training and inference of AI models, in case they are misaligned or are being misused~\cite{miri2024update, un2024governing}. While open-weight models do make this more difficult, their release is not equivalent to a large-scale diffusion of capabilities. This is because in order to run them at scale, one still needs access to specialised compute that is typically housed in centralised clusters and costs more than what most actors can afford. This compute access provides a concrete point of intervention for AI governance. In contrast, by creating a true P2P network as described in Sec.~\ref{sec:p2p}, it might be possible to train or run inference on models in a fully decentralised manner and with no obvious point of accountability or intervention. This problem is further exacerbated in the context of agentic systems~\cite{thornley_shutdown}. While there are important differences between decentralised training and inference, work on the latter is also actively underway~\cite{prime2025inference}. The `no off-switch' problem has been recognised within the decentralised AI community, albeit without any clear solution for now~\cite{long2024protocol}.

\subsection{Compute Structuring}
The methods underpinning both distributed and decentralised training could increase the risk of compute structuring as way of avoiding government oversight. Compute thresholds based on the amount of pre-training FLOPS have been proposed as part of several regulatory frameworks~\cite{euaia_art51}. They rely on the assumption that since compute amounts serve as a rough proxy for the capabilities of the model, they can also indirectly serve as a way of targeting policy interventions towards the riskiest models. To evade such interventions, a malicious actor could divide their training run into several smaller workloads that are executed in parallel using different cloud providers. Such `compute structuring' utilises data parallelism and so greatly benefits from low-communication methods that reduce synchronisation frequency. Detecting parallel structuring might be possible by monitoring the data volume and IP addresses that a compute pool communicates with, but will be challenging without information sharing between various cloud providers~\cite{egan2023oversight}.

Unfortunately, work on KYC schemes and workload monitoring is not progressing at an adequate pace to mitigate this risk. Moreover, strategies to mask communication patterns are likely possible. For example, during the training of INTELLECT-1~\cite{jaghouar2024intellect} over the Internet, synchronisation occurred only once every 38 minutes, and it is possible to purposefully prolong it by up to several training steps, thereby obfuscating any characteristic patterns. Overall, the advent of low-communication training algorithms makes parallel structuring significantly more feasible than before, even if we are uncertain as to how likely it is to be attempted in practice. For concrete proposals on preventing it, see~\cite{comp_struct}.


\subsection{Relevance of Compute}
Our analysis should not be interpreted as suggesting that the importance of compute quantity and quality has been rendered obsolete. Low-communication training changes the way compute can be leveraged, but does not imply that less of it is needed. While more actors may now be able to perform advanced AI training, front-runners in this field will still benefit from vast numbers of GPUs. 

Even after accounting for decentralised training, it is likely that hyperscalers will continue to lead in developing and deploying the most advanced AI models. Any innovation that enhances decentralised training can typically be swiftly adopted by these actors if it offers advantages over their existing methods. In addition, hyperscalers might be able to preserve their competitive moat due to other constraints that continue to bottleneck training and advantage actors with deeper pockets. For instance, ultra-fast GPU memory -- often accounting for nearly half the cost of the most advanced chips -- can make such hardware prohibitively expensive for most~\cite{hbmthenextplatform}. Dominance in compute, whether at a company or country level, will continue to offer significant benefits to AI leadership. As such, we do not expect the relevance of compute (or interventions such as export controls on compute) to be diminished, at least in the foreseeable future.

\subsection{Benefits of Decentralisation}
Democratisation of access to AI systems could be a powerful way of avoiding economic and political disempowerment of individuals and nations~\cite{kulveit2025gradual}. Some argue that in order to achieve this goal, we need a widespread diffusion of capabilities and cannot rely on centralised AI providers~\cite{intelligence_curse}. Cheap, collectively-owned models could provide a solution.

Decentralisation could also significantly advance AI progress by unlocking access to privately held data. While current models primarily utilise public information, vastly larger amounts of private data exist -- such as conversations, car telemetry or CCTV recordings -- that owners are hesitant to share with traditional AI developers. However, privacy-preserving ways of training already exist, and could be supported by decentralised approaches to leverage data without ever transferring it to a central entity. This could lead to models that are more personalised, better reflect local contexts and are tailored to novel applications~\cite{sani2024future, vana2025flower}. This is an area where decentralised training may have a comparative advantage over traditional approaches. Moreover, since data does not need to cross jurisdictions, this could also support international cooperation on AI efforts while maintaining data sovereignty.

\section{Conclusions}
In this paper, we have presented recent developments in the domain of low-communication training algorithms and described two new paradigms that emerge as a result. We argued for making the distinction between multi-data centre \textit{distributed} training and \textit{decentralised} training, which will enable more precise discourse. We discussed several implications of these developments, from possible capability proliferation to the `no off-switch' problem, but also the potential creation of more accessible, tailored AIs.

There are several ways to take this work forward:
\begin{itemize}
    \item Creating robust methods of detecting parallel compute structuring rises in importance. This could be done through a combination of KYC schemes, workload monitoring and on-chip mechanisms, but work in this field is preliminary and should be accelerated~\cite{comp_struct}.
    \item It is unclear to what extent low-communication training can keep up with the frontier of traditional training performed by hyperscalers. Open questions remain about whether improvements similar to DiLoCo can be achieved in other types of parallelism, such as tensor or pipeline parallelisms. It would be especially useful to adapt existing simulation tools to include low-communication parallelism of all types, even if they remain speculative for now~\cite{epoch_simulator, douillard2025streaming, exo_gym}.
    \item Likewise, governance discourse would benefit from a precise estimate of computing power that could be pooled together from consumer GPUs, as well as data centre GPUs in small- to medium-sized clusters. This would enable projecting the scale of decentralised training runs more precisely.
    \item In terms of regulatory interventions such as export controls, it may be useful to include a focus on intra-node networking equipment, as well as GPU memory bandwidth and latency. These could offer more precise control over the use of chips for training at scale.
\end{itemize}
\section*{Impact Statement}
This paper presents technical insights into Machine Learning to enable better AI governance and policymaking. We hope it contributes to a balanced discussion of the risks and benefits of powerful AI systems.

\section*{Acknowledgments}
We would like to thank Matthew Wearden, Kevin Wei, Nathan Helm-Burger, Zilan Qian, as well as anonymous ICML reviewers for helpful comments and discussions. JK graciously acknowledges the financial support of Open Philanthropy. 


\bibliography{example_paper}
\bibliographystyle{icml2025}

\newpage

\appendix
\section{Low-Communication Data-Parallel Training} \label{app:low-com}
\subsection*{Data Parallelism}
\begin{figure}
    \centering
    \begin{tikzpicture}[
  box/.style = {rectangle, draw=black, rounded corners, minimum width=1.8cm, minimum height=1cm, align=center},
  server/.style = {rectangle, draw=black, rounded corners, fill=blue!20, minimum width=3.5cm, minimum height=1cm},
  replica/.style = {rectangle, draw=black, rounded corners, fill=orange!30, minimum width=2.0cm, minimum height=1cm}
]

    \node[server] (server) {Parameter Server};
    
    \node[replica, below=1.5cm of server] (r2) {Worker 2};
    \node[replica, left=1cm of r2] (r1) {Worker 1};
    \node[replica, right=1cm of r2] (r3) {Worker N};
    \node[right=0.1cm of r2] (dots) {\Large$\ldots$};
    \foreach \x in {r1, r2, r3}
      \draw[->, thick] (\x) -- (server);
    \foreach \x in {r1, r2, r3}
      \draw[->, thick] (server) -- (\x);
    
    \end{tikzpicture}
    \caption{In data-parallel training, there are $N$ workers which train local replicas of the model on their assigned data shards. After updating the weights, these changes need to be synchronised across all workers. In the most basic implementation, this can be done through a single `parameter server', but more efficient strategies are used in practice.}
    \label{fig:data-parallel}
\end{figure}
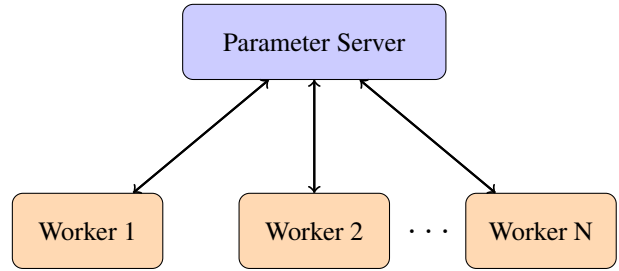

In the context of LLM training, data parallelism refers to the strategy of distributing a model’s training workload across multiple processing units. Each unit -- typically a GPU or a group of GPUs -- holds a full copy of the model and processes a distinct `shard' (a subpart) of the dataset. After computing local gradient updates based on its assigned shard, each unit shares these gradients (or the updated weights) with the others. The updates are then averaged to maintain consistency across the model replicas (see Fig.~\ref{fig:data-parallel}).

Traditionally, this synchronisation step occurs after every training batch, which guarantees strict consistency, but imposes significant communication costs. For example, if processing a batch of tokens takes 4 seconds, while synchronisation requires an additional second, then 20\% of total training time is lost to communication. In large-scale settings, where models might occupy hundreds of gigabytes, this becomes a bottleneck even within a single data centre -- and entirely prohibitive across multiple data centres.

\subsection*{Challenges of High-Frequency, High-Bandwidth Synchronisation}
Standard data parallelism assumes fast, stable interconnects like NVLink or InfiniBand. Extending it to multi-cluster settings quickly becomes unviable due to synchronisation frequency and peak bandwidth requirements. Even if the connection is idle most of the time, it must support extremely high peak throughput when communication occurs. To illustrate the problem, suppose model synchronisation takes 60 seconds over the public Internet (30 seconds both ways), and local training still takes 4 seconds per batch. If synchronisation occurs after every batch, compute utilisation drops to just 6.25\%. At this point, most of the GPU’s time is spent waiting for data rather than training the model. Reducing the synchronisation frequency alleviates this problem, but naive approaches severely degrade training performance due to diverging model replicas and outdated gradients.

\subsection*{Emergence of Low-Communication Training Methods}
Recent algorithmic work -- most notably DiLoCo (Distributed Low Communication), DeMo (Decoupled Momentum), and their successors -- represents a significant breakthrough~\cite{douillard2023diloco, liu2024asynchronous, douillard2025streaming, kale2025eager, peng2024decoupled}. These methods either modify the training protocol to allow workers to synchronise infrequently (as little as once every 500 steps), or with the same frequency, yet transferring a much smaller volume of data. 

For example, the core idea of DiLoCo is to detach local optimisation from global synchronisation through the use of two separate optimisers. One optimiser (often a variant of stochastic gradient descent) handles the local model updates, while a second one adjusts the global parameters after periodic synchronisation. This design maintains stability and convergence even when dealing with outdated gradients. Early empirical studies indicate that such low-communication methods achieve comparable or even better performance than their traditional counterparts.

This training protocol leads to very large gains in compute utilisation. Using the previous example of training over the Internet, synchronising model replicas every 500 steps ($\raisedsim$33 minutes) instead of every step improves compute utilisation from 6.25\% to over 97\%. In practical terms, this enables meaningful training runs over typical Internet connections -- a key requirement of decentralised AI.

\subsection*{Additional Benefits and Follow-Up Improvements}
Apart from reducing communication needs, DiLoCo-style algorithms have other important benefits. Empirical work has demonstrated several properties that make them very well-suited for decentralised and multi-data centre training:
\begin{itemize}
    \item \textbf{Asynchronous communication}: DiLoCo-style optimisers are robust to asynchronous averaging, that is situations where only a subset of replicas can average their parameters at a given point. This property is useful from the perspective of training over unstable connections, where individual contributors may drop in and out of the compute pool.

    \item \textbf{Robustness to data shards from different distributions}: performance does not degrade when workers train on datasets that follow distinct underlying distributions (for instance, with varying numbers of training samples related to a certain feature). This is key to unlocking vast amounts of privately-held data, which naturally will have a high degree of variation between individual contributors.

    \item \textbf{Improved performance with model scale}: perhaps most importantly, follow-up work by~\cite{charles2025communication} showed that in certain settings, DiLoCo and its successors can \textit{outperform} traditional data-parallel training altogether -- with improvements in both convergence speed and final loss. Their analysis also found that the effectiveness of these methods improves with model size, suggesting that low-communication training might be able to keep up with the scaling paradigm.
\end{itemize}

Overall, these properties are in stark contrast to normal data parallelism, which is typically fragile to node failures, requires homogeneous hardware, and depends on continuous high-speed connectivity.

\section{Reasoning Models and Decentralised Compute} \label{app:reasoning}
Since late 2024, there has been an emergence of so-called reasoning models, such as OpenAI's `o' series, Claude 3.7 Sonnet-thinking, and DeepSeek-R1. These models are deliberately taught to `think' before producing the final answer, typically generating hundreds or thousands of tokens during an interaction. This implies a rebalancing in AI workloads, since serving the model to customers now requires a growing amount of compute -- a phenomenon often referred to as `inference-time scaling'. Crucially, it also bears important implications for their training process. Below, we explain why reasoning models can more easily accommodate low-bandwidth environments during post-training than traditional models, making them more compatible with distributed and decentralised setups.

\subsection*{Reasoning Models Are Suited for Low Communication Environments}
Most reasoning models are post-trained using reinforcement learning (RL), which differs in several fundamental ways from the self-supervised learning (SSL) used in pre-training. SSL involves generating predictions for a known sequence of tokens (known as a `forward pass'), comparing those predictions to the ground truth, and adjusting the model weights such that the next prediction is incrementally better (`backward pass'). Importantly, this process involves a 1:1 ratio of forward and backward passes -- every forward pass must be accompanied by a weight update. As discussed in Appendix~\ref{app:low-com}, synchronising these updates across all nodes in a network is highly communication-intensive and generally requires fast, stable connectivity.

In contrast, post-training with RL involves a much larger number of forward passes per gradient update. The model explores a large space of possible `thinking trajectories', each consisting of many new tokens generated one at a time. These trajectories are scored using a reward function, and only then used to perform an update to the model weights. In this setup, the ratio of forward to backward passes can easily reach 1000:1. This naturally reduces the frequency of weight updates, which in turn reduces the frequency of synchronisation. In settings with limited bandwidth, such as those encountered in distributed or decentralised training, this is a highly desirable property.

\subsection*{Practical Implications}
The shift to reasoning models lowers the bar for contributing compute to advanced AI model training. Normally, in order to take part in a training run, we need to have enough compute to fit the model into memory, calculate the gradients in a backward pass, and synchronise them with other nodes in the network. However, generating thinking trajectories for RL-based post-training is less compute intensive and requires less memory, opening the door to the usage of consumer-grade hardware. To give a concrete example, a single M3 Ultra platform by Apple, which comes with up to 512 gigabytes of memory, is sufficient to run inference on models as large as DeepSeek-R1 (671 billion parameters). Moreover, such nodes do not need to be connected to their peers with high-bandwidth networks, as they only have to communicate the generated thinking trajectories and their corresponding rewards, not the full gradients. Thus, this aspect of RL training can be distributed (or even decentralised) across thousands of dispersed contributors, each of whom can utilise typical Internet connections.


\end{document}